\pdfoutput=1
\documentclass[aps, prb, twocolumn, 10pt, floatfix]{revtex4-1}
\usepackage{hyperref}
\usepackage{graphicx}
\usepackage{amsmath}
\usepackage[caption=false]{subfig}

\usepackage{amssymb}

\usepackage{bm}										
\usepackage{braket}

\begin{document}

\title{ Simulating open quantum dynamics with time-dependent variational matrix product states: Towards microscopic correlation of environment dynamics and reduced system evolution  }
\author{Florian A. Y. N. Schr\"oder}
\email{fayns2@cam.ac.uk}

\author{Alex W. Chin}
\affiliation{Cavendish Laboratory, University of Cambridge, J. J. Thomson Avenue, Cambridge, CB3 0HE, UK}

\begin{abstract}
	We report the development of an efficient many-body algorithm for simulating open quantum system dynamics that utilizes a time-dependent variational principle for matrix product states to evolve large system-environment states. Capturing all system-environment correlations, we reproduce the non-perturbative, quantum-critical dynamics of the zero temperature spin-boson model, and then exploit the many-body information to visualize the complete time-frequency spectrum of the environmental excitations. Our 'environmental spectra' reveal correlated vibrational motion in polaronic modes which preserve their vibrational coherence during incoherent spin relaxation, demonstrating how environment information could yield valuable insights into complex quantum dissipative processes.
\end{abstract}

\maketitle
\section{Introduction}\label{sec:intro}
Dissipative quantum dynamics can now be probed in microscopic, real-time detail, yielding unprecedented insight into system-environment processes whose understanding and control will be essential for future quantum technologies \cite{collini2010coherently,engel2007evidence,ferraro2014real,hase2003birth,romero2014quantum,shulman2014suppressing}.
Recently, experimental observations of coherence in organic and biological materials \cite{collini2010coherently,engel2007evidence,falke2014coherent,gelinas2014ultrafast,romero2014quantum,bakulin2015vibrational,musser2015evidence}, have strongly motivated a better understanding of the \emph{microscopic} origin and role of 'ultrafast' ($<$ps) effects resulting from quantum correlations, memory, bath structure and non-perturbative system-bath couplings.
While advanced reduced density matrix techniques can account for these phenomena \cite{iles2014environmental,ishizaki2009unified,Kast2013,Nalbach2010}, information is inevitably discarded when tracing out the environment, and a truly many-body approach is required for deeper insight into the mechanisms at play.
However, this necessitates the evolution of a macroscopically large system-environment state, and the determination of open-system ground states and dynamics are only tractable with powerful computational techniques, such as exact diagonalization, multi-configurational Hartree-Fock and various time-dependent (density matrix) renormalization group techniques \cite{Alvermann2009,Bulla2003,Guo2012,Orth2010,prior2010efficient,wang2003multilayer}.

In this article, we present a versatile new approach to this problem, based on the recently proposed time-dependent variational principle (TDVP) for \emph{variational matrix product states} (VMPS) \cite{Haegeman2014,Lubich2014,Schollwoeck2011,weichselbaum2009variational,Guo2012,frenzel2013matrix}.
This many-body method gathers together several recent advances in VMPS theory to create a fast, efficient algorithm for system-bath dynamics, where resources can be allocated 'on the fly' \cite{Guo2012,Haegeman2014,Lubich2014}. 
We show that the method can correctly capture the complex, non-Markovian physics of the famous spin-boson model (SBM) \cite{Leggett1987,weiss1999quantum}, and then show how visualizing the accompanying environmental dynamics provides an informative, time and frequency-resolved \emph{spectroscopy} of open systems.
This combination of accurate system dynamics and powerful diagnostic tools for analyzing the  \emph{detail} within system-environment states can be applied to a wide range of problems, and could be particularly useful for unraveling the physics of the 'intermediate' regime of open systems \cite{mohseni2014quantum}.

The paper is organized as follows. 
Section \ref{sec:sbm} defines the model Hamiltonian used in our calculations. 
Section \ref{sec:orthpol} and \ref{sec:vmps} briefly outline the orthogonal polynomial chain mapping as well as the VMPS method upon which our algorithm is based, including the optimized boson basis (OBB) which modifies the matrix product state (MPS) network. 
Section \ref{sec:tdvp} presents the new TDVP scheme required to time-evolve the modified MPS and presents a derivation of the scheme.
Finally section \ref{sec:dynamics} presents numerical results for the model Hamiltonian and demonstrates to which detail the environmental dynamics can be analyzed and related to system-bath dynamics.

\section{The spin-boson model}\label{sec:sbm}
The spin-boson model (SBM) has become the benchmark for testing advanced open system methods, as well as having numerous direct applications in physics, chemistry and biology  \cite{Leggett1987,weiss1999quantum}.
It describes a quantum two-level system (TLS) that interacts with an environment of harmonic oscillators via the Hamiltonian ($\hbar = 1$) \cite{Leggett1987},
\begin{equation}
	H=\underbrace{-\frac{\Delta}{2}\sigma_x-\frac{\epsilon}{2}\sigma_z}_{H_S}+\frac{\sigma_z}{2} \sum_n{\lambda_n\left(b_n+b_n^\dagger\right)}+\sum_n{\omega_n b_n^\dagger b_n},
	\label{SBM}
\end{equation}
where the TLS has an energy bias $\epsilon$, coherent tunneling amplitude $\Delta$ and coupling $\lambda_n$ to environmental modes of energy $\omega_n$.
The operators $\sigma_i$ are Pauli matrices, while $b_n^\dagger$ and $b_n$ are bosonic creation and annihilation operators.
Here, we focus on general power-law spectral functions $J(\omega)=\pi\sum_n{\lambda_n^2 \delta\left(\omega-\omega_n\right)}=2\pi\alpha\omega_c^{1-s}\omega^s\theta\left(\omega_c-\omega\right)$ which parameterize the bandwidth of the environment $\omega_c$, the (dimensionless) interaction strength $\alpha$ and the frequency dependence exponent $s$ which defines sub-Ohmic ($s<1$), Ohmic ($s=1$) and super-Ohmic ($s>1$) environments.
The Ohmic and sub-Ohmic cases possess a range of quantum phase transitions (QPT) ($\langle \sigma_{z}\rangle_{g.s}\neq0$) when $\alpha$ exceeds a critical coupling $\alpha_{c}$ \cite{Alvermann2009,bera2014generalized,bera2014stabilizing,Bulla2003,Chin2011,Guo2012, bruognolo2014two, Kast2013,Nalbach2010, Orth2010,Vojta2005,Yao2013, Zhang2010, anders2005real, le2007entanglement,winter2009quantum}.

Since the VMPS method works especially well on $1D$ chain Hamiltonians, it is necessary to transform the SBM.
We use the orthogonal polynomial mapping introduced in Prior \textit{et al.} \cite{Chin2010,prior2010efficient} to obtain a semi-infinite $1D$ coupled-chain representation (see next section and Ref. \cite{Bulla2003}),
\begin{equation}
	\begin{split}
		H=H_S &+ \frac{\sigma_z}{2}c_0\left(a_0+a_0^\dagger\right)\\
		&+\sum_{k=0}^{L-2}\left[\omega_k a_k^\dagger a_k +t_k\left(a_k^\dagger a_{k+1} + a_{k+1}^\dagger a_k \right)\right],
	\end{split}
	\label{eq:chain}
\end{equation}
with effective system-environment coupling $c_0=\sqrt{\int_0^{\omega_c}\! \frac{J(\omega)}{\pi} \, \mathrm{d}\omega}$ and truncation of the chain to a length $L$.

\section{Orthogonal polynomials}\label{sec:orthpol}
The mapping of the environment to a semi-infinite chain model is performed using orthogonal polynomials as described in Ref. \cite{Chin2010, Woods2014}. The star-like Hamiltonian
\begin{equation}
	\begin{split}
		H=H_{S}&+\int\limits_{0}^{x_{max}}\!g(x)b_{x}^{\dagger}b_{x}\, \mathrm{d}x\\
		&+\hat A\int\limits_{0}^{x_{max}}\!h(x)\left(b_{x}+b_{x}^{\dagger}\right)\, \mathrm{d}x,
	\end{split}
\end{equation}
with the system Hamiltonian $H_S$ and the system operator $\hat A$ in the interaction term is mapped onto the chain Hamiltonian
\begin{equation}
	\begin{split}
		H=H_{S}&+\hat A c_{0}\left(a_{0}+a_{0}^{\dagger}\right)\\
		&+\sum_{k=0}^{\infty}\left[\omega_{k}a_{k}^{\dagger}a_{k}+t_{k}\left(a_{k}^{\dagger}a_{k+1} +a_{k+1}^{\dagger}a_{k}\right)\right].
	\end{split}
\end{equation}

For an arbitrary spectral density $J(\omega)$ the mapping can be obtained by finding the recurrence relation of polynomials orthogonal with respect to the weight function
\begin{equation}
	h^2(x)=J(g(x)) \frac{g'(x)}{\pi},
\end{equation} 
where usually the linear dispersion relation $g(x)=\omega_c x=\omega$ is used without loss of generality. The orthonormal polynomials $\tilde{p}_k(x)$ generating the orthogonal transformation $U_k(x)$ from the continuous variable $x$ onto the discrete chain
\begin{equation}
	\begin{split}
		U_k(x)&= h(x) \tilde{p}_k(x), \\
		a_k^\dagger &= \int\limits_0^{x_{max}}\!U_k(x) b_x^\dagger \,\mathrm{d}x,
	\end{split}
\end{equation}
satisfy the normalization condition
\begin{equation}
	\braket{\tilde{p}_k,\tilde{p}_l}_\mu = \int\limits_{0}^{{x_{max}}}\!{h^{2}}\left(x\right){\tilde{p}_{k}}\left(x\right)\tilde{p}_{l}\left(x\right)\,\mathrm{d}x ={\delta_{k,l}},
\end{equation} 
and relate the monic orthogonal polynomials $\pi_k(x)$ via
\begin{equation}
	\tilde{p}_k(x) = \frac{\pi_k(x)}{\left\|\pi_k(x)\right\|_\mu},
\end{equation} 
where the norm $\left\|p\right\|_\mu = \sqrt{\braket{p,p}_\mu}$ is induced by the inner product $\braket{\cdot,\cdot}_\mu$ under the measure $\mathrm{d}\mu(x) = h^2(x)\mathrm{d}x$.
The chain parameters $c_0,\omega_k,t_k$ are related to the coefficients $\alpha_k,\beta_k$ of the monic recurrence relation
\begin{equation}
	\begin{split}
		{\pi_{k + 1}}\left( x \right) &= \left( {x - {{\rm{\alpha }}_k}} \right){\pi_k}\left( x \right) - {{\rm{\beta }}_k}{{\rm{\pi }}_{k - 1}}\left( x \right),\\
		\omega_k&=\omega_c\alpha_k,\\
		t_k &= \omega_c \sqrt{\beta_{k+1}},\\
		c_0 &= \left\| \pi_0(x)\right\|_\mu,
	\end{split}
\end{equation} 
with the initial polynomials $\pi_{-1}(x)=0$ and $\pi_0(x) = 1$.
For the power-law spectral density with hard cut-off at the characteristic frequency $\omega_c$
\begin{equation}
	J(\omega)=2\pi\alpha \omega_s^{1-s} \omega^s\theta\left(\omega_c -\omega\right),
\end{equation} 
the site energies $\omega_k$ and couplings $c_0,t_k$ can be analytically found as
\begin{equation}
	\begin{split}
		{\omega_{k}}&=\frac{{{\rm {\omega}}_{c}}}{2}\left({1+\frac{{s^{2}}}{{\left({s+2k}\right)\left({2+s+2k}\right)}}}\right),\\
		{t_k} &= \frac{{{{\rm{\omega }}_c}\left( {1 + k} \right)\left( {1 + s + k} \right)}}{{\left( {s + 2 + 2k} \right)\left( {3 + s + 2k} \right)}}\sqrt {\frac{{3 + s + 2k}}{{1 + s + 2k}}} ,
	\end{split}
\end{equation}
for $k=0,1,\ldots, L-2$ if the chain is truncated to length $L$.

This analytic mapping allows a simple inversion $b_x^\dagger = \sum_k U_k(x) a_k^\dagger$ to obtain observables of the original Hamiltonian from the chain observables. In the presented work we used the spin projected displacement
\begin{equation}
	\begin{split}
		f^{\uparrow/\downarrow}_x &= \braket{\frac{\openone \pm \sigma_z}{2} \frac{b_x^\dagger+b_x}{2}}\\
		&= \sum\limits_{k=0}^{L-2} h(x) \tilde{p}_k(x) \Re[\braket{\frac{\openone \pm \sigma_z}{2} a_k^\dagger}],
	\end{split}
\end{equation}
and the occupation of phonon modes
\begin{equation}
	\braket{b^\dagger_x b_x} = \sum\limits_{k,l=0}^{L-2} h^2(x) \tilde{p}_k(x)\tilde{p}_l(x) \braket{a_k^\dagger a_l},
\end{equation}
where the continuous variable $x$ has to be discretized.

While for time-evolution the widely used logarithmic discretization needs an averaging scheme over multiple calculations to minimize discretization errors, this method is exact and gives accurate results in one run. In fact this mapping can be recovered in the limit $\Lambda\to 1$ of infinitely fine discretization \cite{Bulla2003, Zitko2009}.

\section{Variational matrix product state formulation}\label{sec:vmps}
We now outline the formalism and features of our algorithm, the intricate details of which can be found in in Ref. \cite{Schollwoeck2011,Haegeman2014, Guo2012}.

For a 1D lattice of size $L$ with sites $k$ and corresponding local eigenstate basis $\ket{n_k}$ of dimension $d_k$ an arbitrary state of the Hilbert space $\ket{\Psi} \in \mathcal{H}$ can be written as
\begin{equation}
	\ket{\Psi}=\sum_{\{n_k\}=1}^{d_k} {\Psi_{n_1,\ldots,n_L}\ket{n_1,\ldots,n_L}},
\end{equation}
where the sum is done over every possible combination of $n_k$.
Any $\ket{\Psi}$ can be written as a matrix product state (MPS) $\ket{\Psi_{MPS}} \in \mathcal M \subseteq \mathcal H$ via iterative application of singular value decompositions (SVD) on $\Psi_{n_1,\ldots,n_L}$ resulting in the rank-3 tensors $A(k) \in \mathbb{C}^{D_{k-1}\times D_k\times d_k}$
\begin{equation}
	\ket{\Psi_{MPS}} = \sum_{\{ n_k \} =1}^{d_k} \bm A^{n_1}\bm A^{n_2}\cdots \bm A^{n_L}  \ket{n_1,\ldots,n_L}
\end{equation}
where we have open boundary conditions ($a_0 = a_L = 1$), use the indexing $A_{a_{k-1},a_k}^{n_k} = (A(k))_{a_{k-1},a_k,n_k}= (\bm A^{n_k}(k))_{a_{k-1},a_k}$, and allow to omit the site argument for clarity.

Furthermore we employ an optimized boson basis (OBB), as introduced by Guo \textit{et al.} \cite{Guo2012}, which is realized via an additional map (isometry) $V\in \mathbb C^{d_k\times d_{OBB,k}}$ from the optimized basis $\ket{\tilde n_k}$ into the local basis $\ket{n_k}$
\begin{equation} 
	A_{a_{k-1},a_k}^{n_k} = \sum_{\tilde n_k=1}^{d_{OBB,k}} \tilde A_{a_{k-1},a_k}^{\tilde n_k} V_{n_k, \tilde n_k},
\end{equation}
where we will write $A$ instead of $\tilde A$ throughout this work.
This mapping allows high compression ($d_{OBB,k} \ll d_k$) of the local oscillator basis in case of large variances $\mathrm{Var}\left(n_k\right)$, which has been shown to be highly effective in dealing with quantum critical SBMs \cite{Guo2012}.

\begin{figure}
	\includegraphics{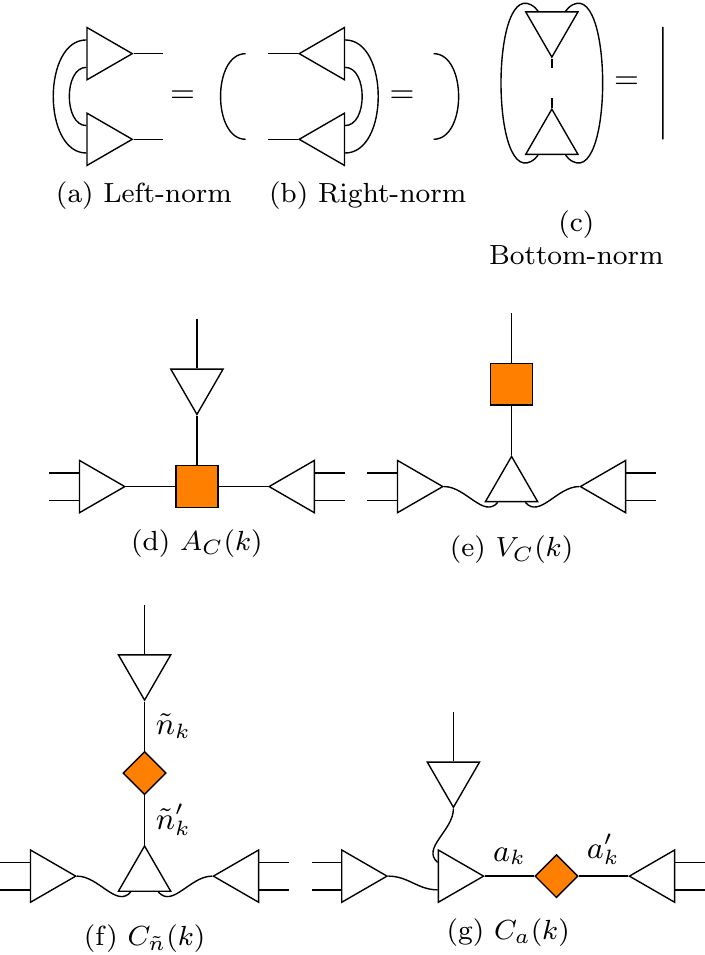}
	\caption{(color online). (a)-(c) All possible normalizations of $A$. The straight line represents the identity matrix. (d)-(g) Center matrices (orange) and corresponding network normalization for one site.\label{fig:si-centermps}}
\end{figure}

Once we represented a state as an MPS, any variational optimization and time-evolution is performed iteratively by sweeping along the chain. During the sweeping procedure we keep the state in a mixed canonical form to benefit from orthogonality conditions.
Furthermore only one matrix can be focused (centered), which means its vectorization $v$ has unit length $\braket{v|v}=1 = \braket{\Psi_{MPS}|\Psi_{MPS}}$.
Fig. \ref{fig:si-centermps} diagrammatically explains all normalizations of $A$ and the possible center matrices $A_C(k), V_C(k)$ and bond centers $C_a(k), C_{\tilde n}(k)$ which have special relevance for the time-evolution explained later.
When focusing on site $k$, all $A$ matrices of site $i<k$ will be kept left-orthonormal while matrices $i>k$ will be right-normalized to produce orthonormal left and right basis states $\ket{\Phi_L},\ket{\Phi_R}$
\begin{multline}
	\ket{\Psi(A)} = \\ \sum_{a_{k-1},n_k,a_k} {\left(A_C(k)\right)^{\tilde{n}_k}_{a_{k-1},a_k} \ket{\Phi_{L,a_{k-1}}^{[1:k-1]}} \ket{\tilde{n}_k} \ket{\Phi_{R,a_{k}}^{[k+1:L]}}}.
\end{multline}

The efficiency of MPS is based on low-rank tensor approximations which significantly reduce the number of variational parameters in the tensors $A(k),V(k)$.
This restricts the Hilbert space spanned by the MPS to a manifold $\mathcal M \subset \mathcal H$.
The combination of high truncation and dynamical 'on the fly' bond adjustment, while representing an optimal manifold allows the implementation of an efficient variational algorithm with significant advantages for computational speed and accuracy \cite{Note1, Schollwoeck2011, weichselbaum2009variational,Guo2012}.
We truncate and expand the bonds $a_k$ and $\tilde n_k$ such that the smallest singular values kept are within $10^{-4}$ and $10^{-4.5}$ leading to adaptive dimensions $D_k$ and $d_{OBB,k}$ with upper bounds $D_{max}$ and $d_{OBB,max}$.

\section{Time-dependent variational principle}\label{sec:tdvp}
The time evolution is performed with the time-dependent variational principle (TDVP) described in Refs. \cite{Haegeman2014, Lubich2014}.
It is derived from the Dirac-Frenkel variational principle and obtained by projecting the Schr\"odinger equation onto the tangent space of the MPS manifold $\mathcal M$ to find optimal equations of motion for each MPS center tensor within $\mathcal{M}$ to generate the best approximation $\ket{\Psi_{MPS}(t)}$ to the exact state $\ket{\Psi (t)}$.
Haegeman \textit{et al.} have shown that this is equivalent to a Lie-Trotter splitting of the MPS instead of the time-evolution operator $\mathcal{U}(t) = e^{-i\hat{H}t}$.
Unlike the Suzuki-Trotter splitting of $\mathcal{U}(t)$, errors only accrue from the integration scheme.

Since the OBB modifies the MPS network, our implementation of the TDVP needed an extension of the original TDVP scheme.
To prove that this can be done and to derive the appropriate integration scheme, we will present in Sec. \ref{sec:tdvp-proof} the derivation of the tangent space projector $\hat{P}_{T_{\ket{\Psi_{MPS}}}\mathcal M}$ in the formalism established by Ref. \cite{Lubich2014}. Additionally we will give a more simplified description of the resulting integration scheme in Sec. \ref{sec:tdvp-scheme}.

\subsection{Derivation of the tangent space projector}\label{sec:tdvp-proof}
This derivation will be similar to Theorem 3.1 of Ref \cite{Lubich2014} but using the left canonical form of the MPS with OBB matrices. To be more consistent with their notation and for clarity, we will use $n_k, \tilde n_k$ and $a_k$ instead of $d_k,d_{OBB,k}$ and $D_k$ to address bond dimensions.

Additionally to the definitions of Lubich \textit{et al.} we need to introduce further notation to address the components of an MPS tensor $X\in \mathbb{R}^{n_1 \times \cdots \times n_L}$ and their unfoldings (as depicted in Fig. \ref{fig:si-centermps}).

Any site tensor $C_k \in \mathbb{R}^{a_{k-1}\times n_k \times a_k}$ can be written as a matrix using the left and right unfolding $\bm C_k^< \in \mathbb{R}^{(a_{k-1},n_k)\times a_k}$ and $ \bm C_k^> \in \mathbb{R}^{(a_{k},n_k)\times a_{k-1}}$. Similarly the orthonormalized site tensors $Q_k^<,Q_k^> \in \mathbb{R}^{a_{k-1}\times n_k \times a_k}$ are unfolded as
\begin{equation}
	\begin{split}
		\bm Q_k^< &\in \mathbb{R}^{(a_{k-1}n_k)\times a_k},\\
		\bm Q_k^> &\in \mathbb{R}^{(a_{k}n_k)\times a_{k-1}},\\
	\end{split}
\end{equation}
and additionally fulfill the normalization conditions $ \bm Q_k^{<T}\bm Q_k^< = \openone_{a_k}$ and $ \bm Q_k^{>T}\bm Q_k^> = \openone_{a_{k-1}} $.

Unfolded segments of the MPS to the left and right of site $k$ are denoted as $\bm X_{\leq k-1} \in \mathbb{R}^{(n_1\ldots n_{k-1})\times a_{k-1}}$ and $\bm X_{\geq k+1} \in \mathbb{R}^{(n_{k+1}\ldots n_L)\times a_{k}}$ respectively. If these parts are left and right orthonormalized we will write $\bm Q_{\leq k-1}$ and $\bm Q_{\geq k+1}$.

This notation can be transferred to the OBB tensors $A_k \in \mathbb{R}^{a_{k-1}\times \tilde n_k \times a_k}$ and $\bm V_k \in \mathbb{R}^{n_k \times \tilde n_k}$ forming a decomposition of the site tensors $C_k$ as
\begin{equation}
	\bm C_k^< = \left( \bm V_k \otimes \openone_{a_{k-1}} \right)\bm A_k^<.
\end{equation}
The OBB tensors can also be orthonormalized, denoted as $Q_{A,k}$ and $\bm Q_{V,k} \in \mathbb{R}^{n_k\times \tilde n_k}$, where $\bm Q_{V,k}^T \bm Q_{V,k} = \bm Q_{A,k}^{<,T} \bm Q_{A,k}^< = \openone_{\tilde n_k} $. They compose $ \bm Q_k^< $ and $ \bm Q_k^> $ as:
\begin{equation}
	\begin{split}
		\bm Q_k^< &= \left( \bm Q_{V,k} \otimes \openone_{a_{k-1}} \right) \bm Q_{A,k}^<,\\
		\bm Q_k^> &= \left( \bm Q_{V,k} \otimes \openone_{a_{k}} \right) \bm Q_{A,k}^>.
	\end{split}
\end{equation}
Furthermore it is necessary to define an new unfolding $\check{\bm Q}_k^< \in \mathbb{R}^{(a_{k-1}a_k)\times \tilde n_k}$ to support the following notation.

Let $X^{T_k}$ be the $k$th tensor transpose cycling the $k$th tensor dimension of $X$ to the 1st position while keeping the order of all other dimensions. This permutation can be written as
\begin{equation}
	X(i_1,i_2,\ldots,i_L) = X^{T_k}(i_{\sigma_k(1)},i_{\sigma_k(2)},\ldots,i_{\sigma_k(L)}).
\end{equation}
with the $k$-cycle $\sigma_k = \left(k\quad 1 \quad 2 \quad \cdots \quad k-1 \right) \in \Sigma_L$. The inverse of this transpose is $X^{T_{-k}}$ defined by the $k$-cycle $\sigma_{-k} =\sigma^{-1}_k = \left(k\quad k-1 \quad \cdots \quad 2 \quad 1 \right)$.
The $k$th unfolding $\bm X^{(k)}$ of $X$ denoted with round braces is then equivalent to the 1st unfolding of the $k$th tensor transpose of X
\begin{equation}
	\begin{split}
		\bm X^{(k)} &= \bm X^{T_k \braket{1}} \in \mathbb R^{n_k \times (n_1\ldots n_{k-1}n_{k+1}\ldots n_L)},\\
		X &= \text{Ten}_{(k)}\left[ \bm X^{(k)} \right]=\text{Ten}_1\left[X^{T_k \braket{1}}\right]^{T_{-k}},
	\end{split}
\end{equation}
with the corresponding tensor reconstruction as its inverse.

A matrix $A\in \mathbb{R}^{m\times n}, m>n$ with $\text{rank}(A) = n$ has a left-inverse
\begin{equation}
	\begin{split}
		A^{-1}   &= \left( A^T A \right)^{-1} A^T,\\
		A^{-1} A &= \openone_n,
	\end{split}
\end{equation}
and a projector $P_A$ onto the range of $A$
\begin{equation}
	P_A = A A^{-1},
\end{equation}
which will be used for the definition of the tangent space projector.

The tangent space $T_X\mathcal M$ of any state $X \in \mathcal M$ of the OBB-MPS manifold can be constructed from the orthogonal subspaces $\mathcal V_{A,k},\mathcal V_{V,k}$. We will use the MPS in left canonical gauge including site $k$
\begin{equation}
	\begin{split}
		X = \text{Ten}_k \left[ \left( \bm Q_{V,k} \otimes \bm Q_{\leq k-1} \right) \bm Q_{A,k}^< \bm X_{\geq k+1}^T \right].
	\end{split}
\end{equation}
Thus we have
\begin{widetext}
	\begin{equation}
		\begin{split}
			\mathcal V_{A,k} &= \left\lbrace \text{Ten}_k \left[ \left( \bm Q_{V,k} \otimes \bm Q_{\leq k-1} \right) \delta \bm Q_{A,k}^< \bm X_{\geq k+1}^T \right] : \delta Q_{A,k}^< \in \mathbb{R}^{a_{k-1}\times \tilde n_k \times a_k}, \bm Q_{A,k}^{<T} \delta \bm Q_{A,k}^<  = 0 \text{ for } k\neq L \right\rbrace,\\
			\mathcal V_{V,k} &= \left\lbrace \text{Ten}_{(k)} \left[\delta \bm Q_{V,k} (\check{\bm Q}_{A,k}^< )^T (\bm X_{\geq k+1} \otimes \bm Q_{\leq k-1})^T\right] : \delta \bm Q_{V,k}^< \in \mathbb{R}^{n_k \times \tilde n_k}, \bm Q_{V,k}^{T} \delta \bm Q_{V,k}  = 0\right\rbrace,
		\end{split}
		\label{eq:defx}
	\end{equation}
\end{widetext}
with $\bm X^T{\geq L+1} = 1$ since $a_L = 1$. Since these subspaces are mutually disjoint, the tangent space is
\begin{equation}
	T_X\mathcal M = \bigoplus_{k=1}^L \left( \mathcal V_{A,k} \oplus \mathcal V_{V,k}  \right),
\end{equation}
and allows the decomposition of any tangent vector $\delta X \in T_X\mathcal M$
\begin{equation}
	\begin{split}
		\delta X &= \sum_{k=1}^L \left( \delta X_{A,k} + \delta X_{V,k} \right),\\
		\braket{\delta X_{i,k}, \delta X_{j,l}} &= 0 \quad \text{ for } i\neq j, k\neq l.
	\end{split}
\end{equation}

We want to find the tangent space projector $P_{T_X\mathcal M}$ for an arbitrary $Z\in \mathbb{R}^{n_1\times \ldots\times n_L}$ such that the projection $\delta U = P_{T_X\mathcal M}(Z) \in T_X\mathcal M$ is orthogonal and therefore satisfies
\begin{equation}
	\braket{\delta U,\delta X} = \braket{Z, \delta X} \quad \forall \delta X \in T_X\mathcal M.
\end{equation}
Due to the disjoint subspaces $\mathcal V_{i,k}$ we can decompose
\begin{equation}
	\delta U = \sum_{k=1}^L \left( \delta U_{A,k} + \delta U_{V,k} \right),
\end{equation}
with $\delta U_{i,k} \in \mathcal V_{i,k}$ and find
\begin{equation}\label{eq:si-z-subspace-projection}
	\braket{\delta U_{i,k},\delta X_{i,k} } = \braket{Z, \delta X_{i,k}},
\end{equation}
for $i \in \{A,V\}$. The projection is then defined by the $\delta \bm B$ matrices in
\begin{equation}
	\begin{split}
		\delta \bm U_{A,k}^{\braket{k}} &= \left( \bm Q_{V,k} \otimes \bm Q_{\leq k-1} \right) \delta \bm B_{A,k}^< \bm X_{\geq k+1}^T, \\
		\delta \bm U_{V,k}^{(k)} &= \delta \bm B_{V,k} (\check{\bm Q}_{A,k}^< )^T (\bm X_{\geq k+1} \otimes \bm Q_{\leq k-1})^T,
	\end{split}
\end{equation}
with $\bm Q_{A,k}^{<T} \delta \bm B_{A,k}^<  = 0$ for $k\neq L$ and $\bm Q_{V,k}^{T} \delta \bm B_{V,k}  = 0$.

To find expressions for $\delta \bm B_{A,k}^<$ and $\delta \bm B_{V,k}$ we follow the same steps as Ref. \cite{Lubich2014}. Here we give an outline for $\delta \bm B_{V,k}$ to obtain $\delta \bm U_{V,k}^{(k)}$.
First substitute the expressions for $\delta \bm U_{V,k}^{(k)}$ and $\delta \bm X_{V,k}^{(k)}$ (Eq. \ref{eq:defx}) into Eq. \ref{eq:si-z-subspace-projection} and isolate the term $\delta \bm Q_{V,k}$ in both inner products. 
\begin{equation}
	\begin{split}
		\braket{\delta \bm U_{V,k}^{(k)},\delta \bm X_{V,k}^{(k)}} &= \braket{
			\delta \bm B_{V,k} \bm M^T,
			\delta \bm Q_{V,k} \bm M^T
		}\\
		& = \braket{
			\delta \bm B_{V,k} \bm M^T \bm M,
			\delta \bm Q_{V,k}
		},\\
		\braket{\bm Z^{(k)},\delta \bm X_{V,k}^{(k)}} &= 	\braket{
			\bm Z^{(k)} \bm M,
			\delta \bm Q_{V,k}
		},
	\end{split}
\end{equation}
where we defined $\bm M = (\bm X_{\geq k+1} \otimes \bm Q_{\leq k-1}) \check{\bm Q}_{A,k}^<$ for clarity.
To remove the inner products on both sides we use the property $\braket{A,C} = \braket{B,C} \to P_C A = P_C B$ with $\bm P_{\delta \bm Q_{V,k}} = \openone_{n_k} - \bm P_{V,k}$ and then substitute the found expression for $\delta \bm B_{V,k}$ into $\delta \bm U_{V,k}^{(k)}$.
\begin{equation}
	\begin{split}
		\delta \bm B_{V,k} &=
		(\openone_{n_k} - \bm P_{V,k})
		\bm Z^{(k)}
		\bm M (\bm M^T \bm M)^{-1},\\
		\delta \bm U_{V,k}^{(k)} &=
		\left( \openone_{n_k}-\bm P_{V,k} \right)
		\bm Z^{(k)}
		(\bm M \bm M^{-1}),
	\end{split}
\end{equation}
Thus we derived the projection of $Z$ onto the tangent space as
\begin{equation}
	\begin{split}
		\delta \bm U_{A,k}^{\braket{k}} &= \left[ \left( \bm P_{V,k} \otimes \bm P_{\leq k-1} \right)-\bm P_{\leq k} \right]\bm Z^{\braket{k}} \bm P_{\geq k+1}, \\
		\delta \bm U_{V,k}^{(k)} &= \left( \openone_{n_k}-\bm P_{V,k} \right) \bm Z^{(k)} \bm P_{\bm M}.
	\end{split}
\end{equation}
where we used the projectors 
\begin{equation}
	\begin{split}
		\bm P_{V,k} 		 &= \bm Q_{V,k}\bm Q_{V,k}^T,\\
		\bm P_{\leq k-1} &= \bm Q_{\leq k-1}\bm Q_{\leq k-1}^T,\\
		\bm P_{\geq k+1}  &= \bm X_{\geq k+1} \bm X_{\geq k+1}^{-1}=\bm Q_{\geq k+1}\bm Q_{\geq k+1}^T.\\
	\end{split}
\end{equation}

In the notation of Ref. \cite{Haegeman2014} the tangent space projector thus reads:
\begin{equation}
	\begin{split}
		\hat{P}_{T_{\ket{\Psi_{MPS}}}\mathcal M} &= \sum_{k=1}^L {\hat{P}_L^{[1:k-1]} \otimes \hat P_V^{[k]}\otimes\hat{P}_R^{[k+1:L]}}\\
		&-\sum_{k=1}^{L-1} {\hat{P}_L^{[1:k]} \otimes \hat{P}_R^{[k+1:L]}},\\
		&+\sum_{k=1}^L {\hat{P}_{A}^{[1:k-1,k+1:L]} \otimes (\openone_k-\hat P_V^{[k]})}.
	\end{split}
\end{equation} 
with the left and right chain projectors
\begin{equation}
	\begin{split}
		\hat{P}_L^{[1:k]}  &= \sum_{a_k=1}^{D_k}{\ket{\Phi_{L,a_k}^{[1:k]}}\bra{\Phi_{L,a_k}^{[1:k]}}},\\
		\hat{P}_R^{[k+1:L]}&= \sum_{a_k=1}^{D_k}{\ket{\Phi_{R,a_k}^{[k+1:L]}}\bra{\Phi_{R,a_k}^{[k+1:L]}}},\\
		\hat{P}_V^{[k]}&= \sum_{\tilde n_k=1}^{d_{OBB,k}}{\ket{\Phi_{\tilde n_{k}}^{[k]}}\bra{\Phi_{\tilde n_{k}}^{[k]}}},\\
		\hat{P}_A^{[1:k-1,k+1:L]}&= \sum_{\tilde n_k=1}^{d_{OBB,k}}{\ket{\Phi_{A,\tilde n_k}^{[1:k-1,k+1:L]}}\bra{\Phi_{A,\tilde n_k}^{[1:k-1,k+1:L]}}},
	\end{split}
\end{equation}

Using these expressions to project $H\ket{\Psi_{MPS}}$ onto $T_{\ket{\Psi_{MPS}}}\mathcal M$ we get, after a Lie-Trotter decomposition of the MPS, a system of equations of motion for each MPS tensor $V_C(k)$ and $A_C(k)$ as well as for the bond matrices $C_{\tilde n}(k)$ and $C_a(k)$ which can be solved directly by
\begin{align*}
	\bm V_C (k,t) &= e^{-i H_V(k) t} \bm V_C(k,0), \\
	\bm C_{\tilde n} (k,t) &= e^{+i H_{\tilde n}(k) t} \bm C_{\tilde n}(k,0), \\
	\bm A_C (k,t) &= e^{-i H_A(k)  t} \bm A_C(k,0), \\
	\bm C_{a} (k,t) &= e^{+i H_{a}(k) t} \bm C_{a}(k,0).
\end{align*}
The effective local Hamiltonians of each tensor used in the exponentials are obtained by a full contraction of the Hamiltonian with the MPS, while omitting the respective tensor. A diagrammatic representation is given in Fig. \ref{fig:si-effhamilton}.

\subsection{1-tensor update scheme}\label{sec:tdvp-scheme}
\begin{figure}
	\includegraphics{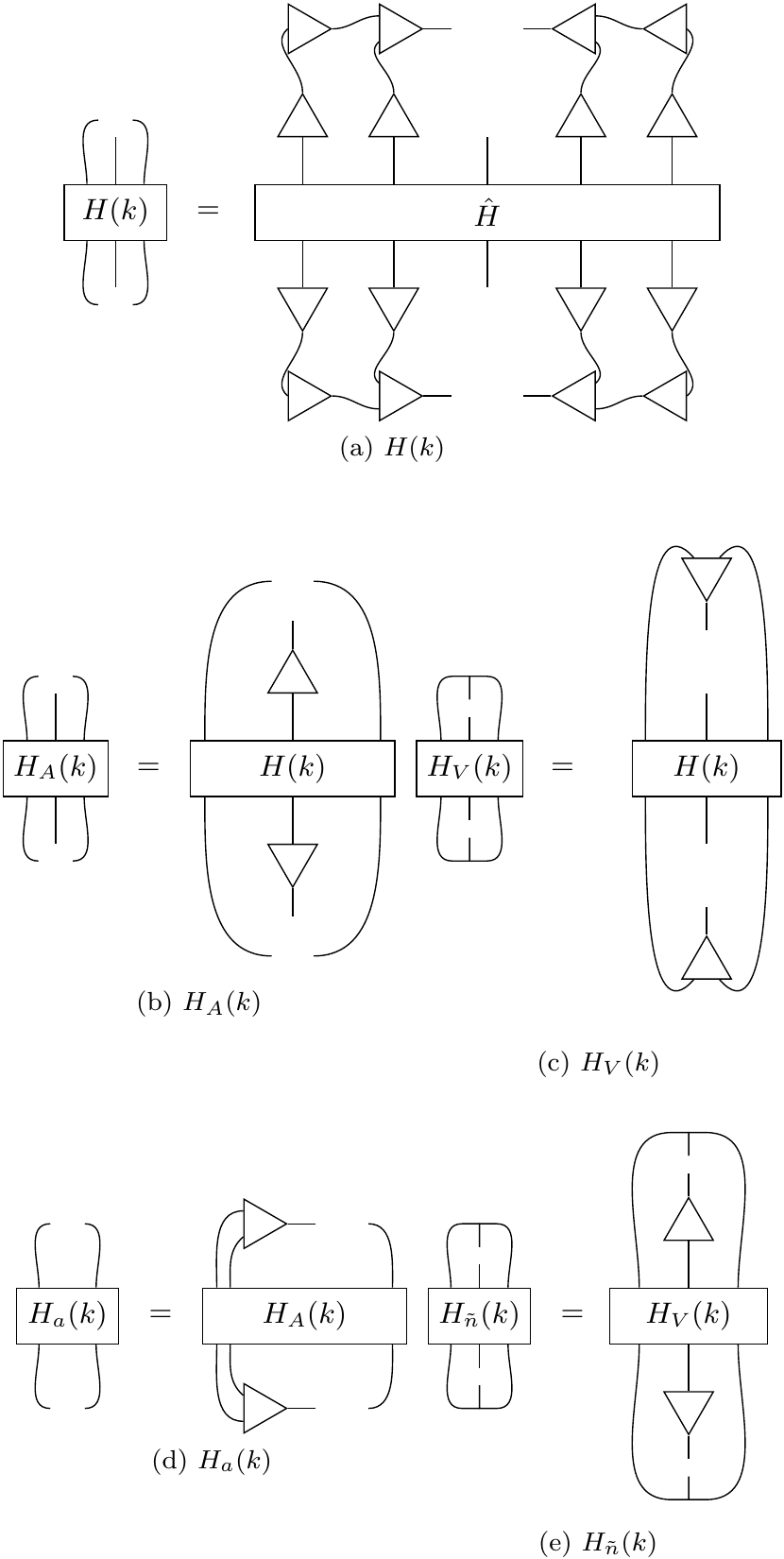}
	\caption{All effective local Hamiltonians needed for the TDVP with OBB.\label{fig:si-effhamilton}}
\end{figure}

A key formal result we present is this new OBB-TDVP scheme with a single-tensor update, which is given by 
\begin{equation}
	\begin{split}
		\bm V_C (k,t+\Delta t) &= e^{-i H_V(k) \Delta t} \bm V_C(k,t), \\
		& \downarrow \text{SVD}\\
		\bm C_{\tilde n} (k,t) &= e^{+i H_{\tilde n}(k) \Delta t} \bm C_{\tilde n}(k,t+\Delta t), \\
		& \downarrow \text{SVD}\\
		\bm A_C (k,t+\Delta t) &= e^{-i H_A(k) \Delta t} \bm A_C(k,t), \\
		& \downarrow \text{SVD}\\
		\bm C_{a} (k,t) &= e^{+i H_{a}(k) \Delta t} \bm C_{a}(k,t+\Delta t), \\
	\end{split}
\end{equation}
where bold notation indicates vectorizations of the tensors. The effective local Hamiltonian $H_{\sim}(k)$ generating the optimal time-evolution of a center matrix is obtained by full contraction with all MPS matrices except for the to be evolved center matrix as depicted in Fig. \ref{fig:si-effhamilton}.
This scheme can be seen to maintain the general form of the TDVP equations given in Refs. \cite{Haegeman2014, Lubich2014}.
Essentially, after each forward evolution $t\to t+\Delta t$ of a centered tensor $V_C$ or $A_C$, the centered bond matrix $C_{\tilde n/a}(t+\Delta t)$ obtained via SVD has to be evolved backwards in time $t+\Delta t \to t$ before contraction with the next tensor.
By sweeping left-to-right and back to the left with a timestep $\Delta t/2$ one obtains a second order symmetric integrator with error of order $\mathcal{O}(\Delta t^3)$.
Since for each evolution step the entire effective Hamiltonian is applied, it is possible to include long-range interactions.

The single-tensor update with OBB has significant advantages over the 1-site update TDVP without OBB in cases where a large number of local states has to be considered. The two additional evolution steps for $\bm V_C$ and $\bm C_{\tilde n}$ can outweigh the unfavorable scaling of the 1-site TDVP without OBB.
The computational complexity of the TDVP with OBB scales as $\mathcal{O}(\max(D^3d_{OBB},D^2d_{OBB}^2,d_{OBB}d^2,d_{OBB}^2d))$ which is significantly faster than the original scaling $\mathcal{O}(\max(D^3d,D^2d^2))$ without OBB for $d_{OBB}\ll d$.
The scaling behavior of OBB-TDVP is also more preferable in comparison with the common time-evolving block decimation (TEBD) ($\mathcal{O}(d^3D^3)$) and the very recently proposed TEBD with local basis optimization (TEBD-LBO) of Brockt \textit{et al.} \cite{Brockt2015} with a scaling of $\mathcal{O}(\max(d_{OBB}^3D^3,d^3D^2))$.
The latter uses a technique very similar to the OBB to map the local Hilbert space onto an optimized basis to reduce the computational effort.

\begin{figure}
	\includegraphics[width=\linewidth]{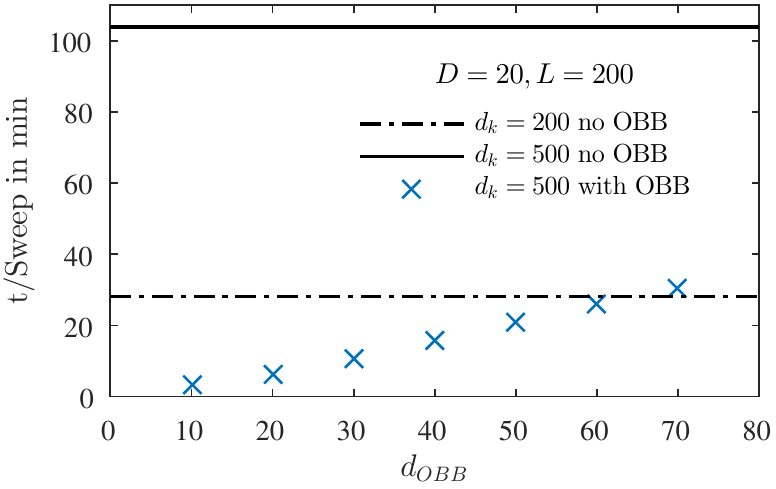}
	\caption{\label{fig:si-TDVP-OBB-comparison}(color online). Comparison of CPU time per sweep for TDVP without OBB and $d_k=200, d_k=500$ (lines) and for TDVP with OBB and $d_k=500$ (crosses).}
\end{figure}

As evident from the example given in Fig. \ref{fig:si-TDVP-OBB-comparison} it is possible to increase the local Hilbert space on each site from $d_k=200$ to $d_k=500$ at no extra cost by introducing an OBB of $d_{OBB}=65$. Similarly a calculation needing $d_k=500$ can be accelerated by a factor of 4-10 by using an OBB, depending on the amount of (allowed) entanglement. Generally, the smaller $d_{OBB}$, the less entanglement can be captured by the MPS between a site and its left and right environment, but the faster and more memory efficient is the computation.

\section{Simulated SBM dynamics}\label{sec:dynamics}
If not otherwise stated, we take $\Delta = 0.1, \epsilon = 0, \omega_c = 1, \Delta t= 0.1-1$ with initial MPS dimensions $d_k =30, D=5, d_{OBB} = 5$ and maximal $D_{max} = 5, d_{OBB,max}=15$ in all simulations.
All results had converged sufficiently w.r.t. $D$ by $D=5$, as shown in the appendix.
The required chain length $L$ depends on $\omega_c$, the simulated time range $T$ and the extracted observables.
If only the system dynamics are desired, we take $L=\frac{2}{7}T$, while for environment observables in frequency space, we need $L=\frac{2}{3.5}T$ to avoid artifacts caused by unphysical reflections at the end of the chain (see Fig. \ref{fig:si-results-chain}).
For reference, a single sweep with dimensions $d_k=30, D=5, d_{OBB}=15$, and $L = 200$ takes 4 seconds on one core of an Intel Core i7-4790 CPU.

We will consider two different initial state preparations for time-evolution, both with the environment at zero temperature.
The polarized coupled state is obtained via variational ground state optimization of the z-polarized spin coupled to the environment, as in Refs. \cite{Orth2010,Kast2013}.
The uncorrelated product state has the vacuum state $\ket{\Omega}$ of the bath coupled to the z-polarized spin.

\subsection{Spin-dynamics}

\begin{figure}
	\includegraphics[width=\linewidth]{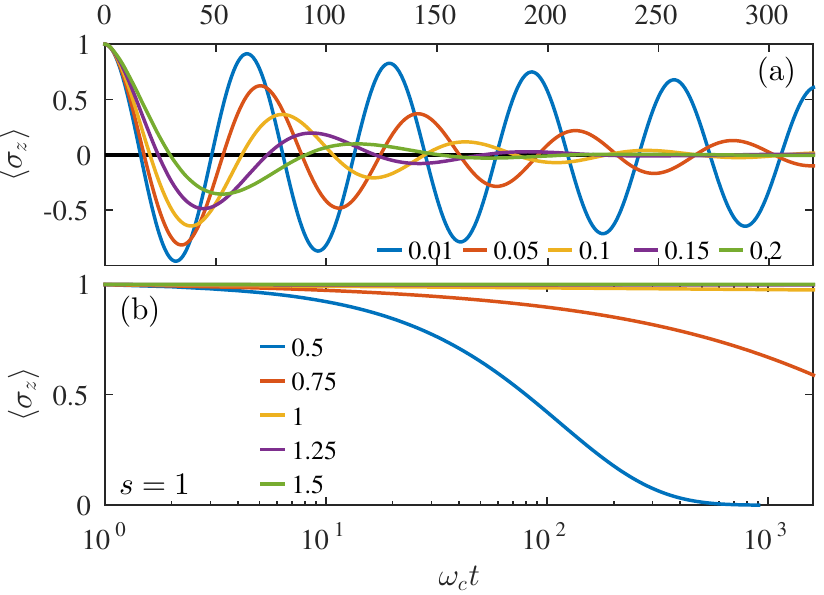}
	\caption{\label{fig:results-sz-dynamics-ohmic}(color online). (a) Weak ($\alpha < \alpha_c$) and (b) strong coupling for Ohmic ($s=1$, $\alpha_c=1$) spectral density at different $\alpha$. (b) was obtained at $\Delta t = 0.01$.}
\end{figure}
\begin{figure}
	\includegraphics{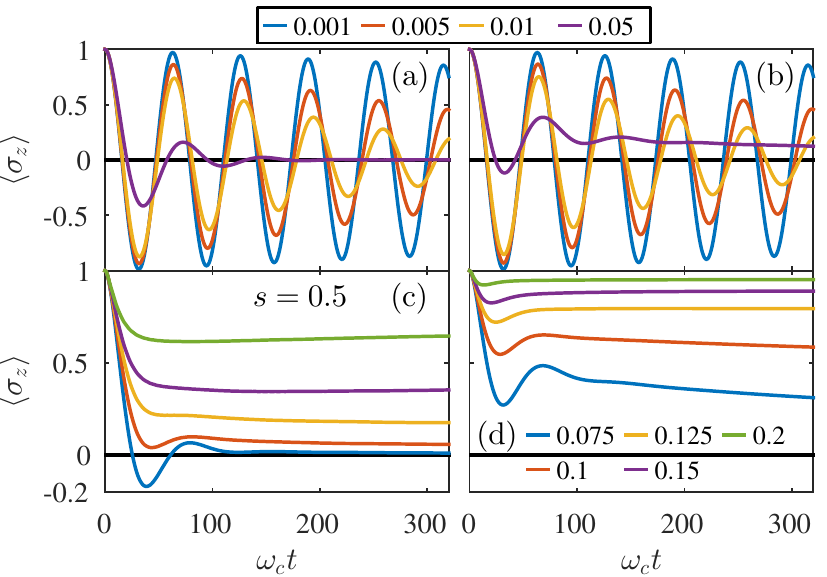}
	\caption{\label{fig:results-sz-dynamics-subohm}(color online). (a)+(b) Weak and (c)+(d) strong coupling (below and above $\alpha_c$) for sub-Ohmic spectral density ($s=0.5$, $\alpha_c = 0.107$) at different couplings $\alpha$. Initial state is (a)+(c) the product state and (b)+(d) the coupled state.}
\end{figure} 
\begin{figure}
	\includegraphics{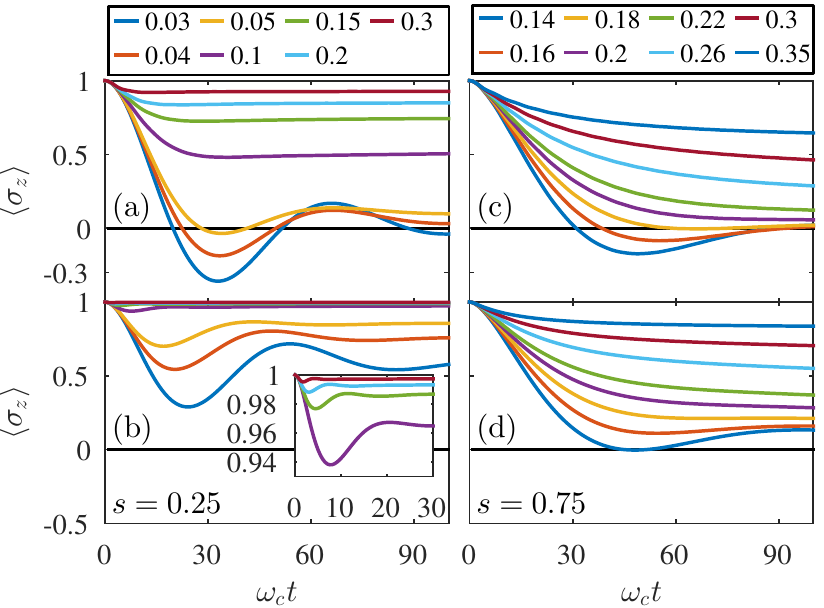}
	\caption{\label{fig:results-sz-dynamics-subohm-kast}(color online). Weak and strong coupling (below and above $\alpha_c$) for sub-Ohmic spectral densities ((a)+(b): $s=0.25$, $\alpha_c = 0.022$; (c)+(d): $s=0.75$, $\alpha_c \approx 0.3$, $\alpha_{CI} \approx 0.22$) at different couplings $\alpha$. Initial state is the product (a)+(c), or the coupled state (b)+(d).}
\end{figure}

In Figs. \ref{fig:results-sz-dynamics-ohmic}-\ref{fig:results-sz-dynamics-subohm-kast} we present the dynamics in the weak ($\alpha < \alpha_c$) and strong coupling regime ($\alpha > \alpha_c$) for Ohmic $s=1$ and sub-Ohmic $s=0.25, 0.5, 0.75$ spectral densities.
Globally, we find excellent agreement with previous numerical work on the non-Markovian dynamics of the SBM, such as the time-dependent numerical renormalization group (TD-NRG) \cite{Orth2010} and path integral methods \cite{Kast2013}.
In brief, the Ohmic results (Fig. \ref{fig:results-sz-dynamics-ohmic}) show increasingly damped oscillations as $\alpha$ increases for $\alpha<0.5$, overdamped relaxation for $0.5<\alpha<1$ which relax more slowly as $\alpha\rightarrow1$ and complete localization above the quantum critical coupling of $\alpha_{c}=1$.

The spin dynamics of the sub-Ohmic SBM close to the coherent-incoherent changeover $\alpha_{CI}(s)$ are shown under the two different initial preparations of product state and coupled state in Fig. \ref{fig:results-sz-dynamics-subohm} and \ref{fig:results-sz-dynamics-subohm-kast}.
In contrast to the Ohmic case, the sub-Ohmic dynamics always remain underdamped for $s=0.25$, showing at least one oscillation even above $\alpha_c=0.022$, as recently found in \cite{Orth2010, Kast2013, Yao2013}.
Furthermore an initial polarization of the bath leads to a higher frequency of spin oscillations with stronger coupling $\alpha$ which is not observable in the product state since the polarization persists on a larger time scale, giving an effective bias to the TLS Hamiltonian \cite{Nalbach2010}.
The frequency of these oscillations is non-monotonic, initially decreasing due to dressing of the tunneling matrix element (a system-bath correlation effect) and then increasing for stronger coupling.
The $s=0.5$ case exhibits overdamping only with the product state for $\alpha > \alpha_{CI} \approx \alpha_c=0.107$, while the coupled state always has initial oscillations even at strong coupling.
The final value of $\langle \sigma_{z}\rangle$ reflects the mean field nature of the QPT in systems with $0<s\leq 0.5$, for which the spin magnetization grows continuously from zero above the critical coupling \cite{Alvermann2009,Chin2011, le2007entanglement, winter2009quantum}.
At $s=0.75$ both the product state and the coupled state lead to overdamped dynamics for $\alpha \gtrapprox \alpha_{CI} \approx 0.22$ and localize for $\alpha > \alpha_c\approx 0.3$.
The accuracy of the calculations across a wide range of spectral densities from weak to strong coupling combined with the efficient use of computational resources demonstrates the versatility of the VMPS implementation.

\subsection{Environment dynamics and spectroscopy}
\begin{figure*}
	\includegraphics[width=\linewidth]{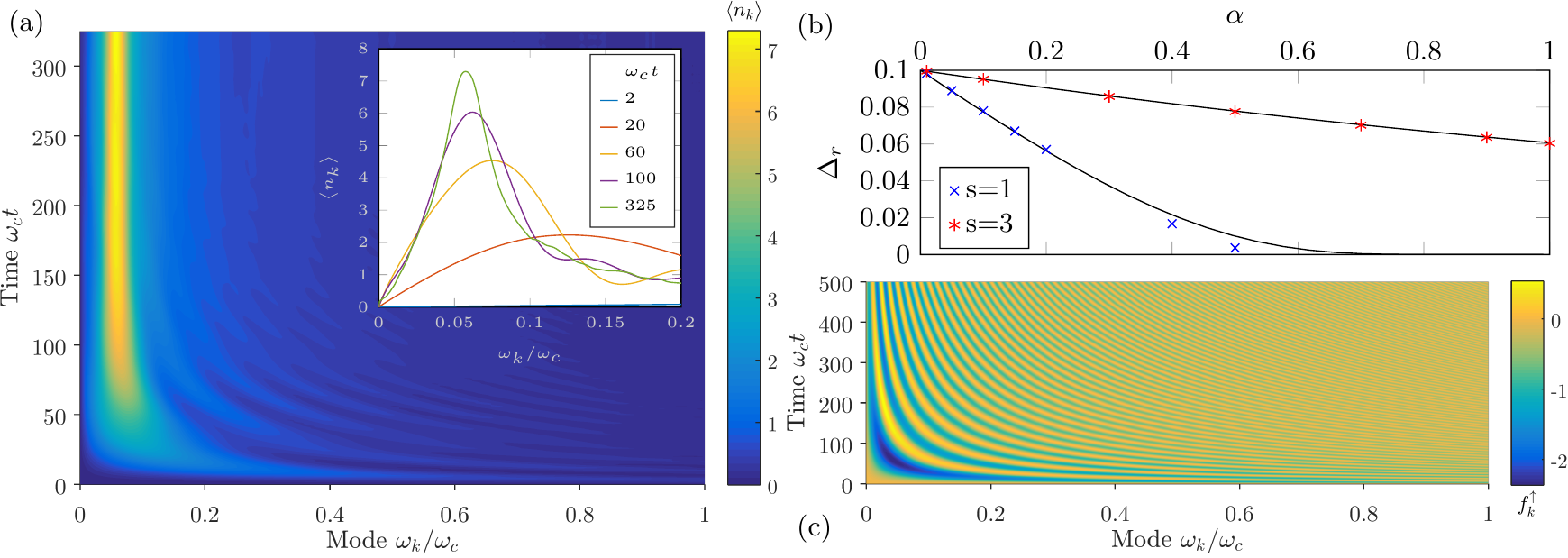}
	\caption{\label{fig:results-bath-omegak}(color online). (a) Phonon modes $\omega_k$ occupation for an Ohmic environment with $\alpha = 0.2$ exhibit a strong resonance at the renormalized tunneling amplitude $\Delta_r \approx 0.057$ and a damping of oscillations. (b) Simulated renormalized tunneling $\Delta_r$ versus analytic expression (lines) shows good agreement. Polaron theory predicts $\Delta_r = \Delta e^{-\frac{\alpha}{s-1}}\text{ for }s=3$. (c) The projected phonon displacement $f_k^\uparrow$ for $s=1, \alpha=0.4$ oscillates even after spin relaxation at $t\approx 350$.}
\end{figure*}

Having verified our method, we now use an efficient inversion of the chain mapping (see Sec. \ref{sec:orthpol}) to present the dynamics of the \emph{entire} environment in time-frequency space \footnote{The dynamics in the chain basis $a_k$ are useful for monitoring the quality of the simulation, i.e. checking for reflections due to the finite length of the environment, and also aid the dynamic reallocation of resources by looking at the entanglement entropy and use of isometries along the chain.
However, as the chain is an artificial expedient for performing the simulations, the original basis $b_n$ gives insight that is easier to relate to analytical theories.}.
Fig. \ref{fig:results-bath-omegak}(a) and its inset shows the population of each mode for intermediate Ohmic coupling, with an initially broad excitation and subsequent emergence of a sharp resonance peak around the TLS energy gap.
The peak in this environmental 'absorption spectrum' rises on the timescale of spin relaxation ($\omega_{c}t\approx 300$), while its position evolves and is complete by $\omega_{c}t\approx 110$.
The final peak position is reached after about one period of the resonant frequencies of the environment modes ($\omega_{res}\approx0.06\omega_{c}$), consistent with the 'sampling' required for the broadband environment to resolve the TLS gap.
A range of novel phenomena related to non-detailed balance and phase-dependent relaxation have recently been predicted to exist prior to this time \cite{2015arXiv150608151O}, though we will not explore this further here.

\begin{figure}
	\includegraphics[width=\columnwidth]{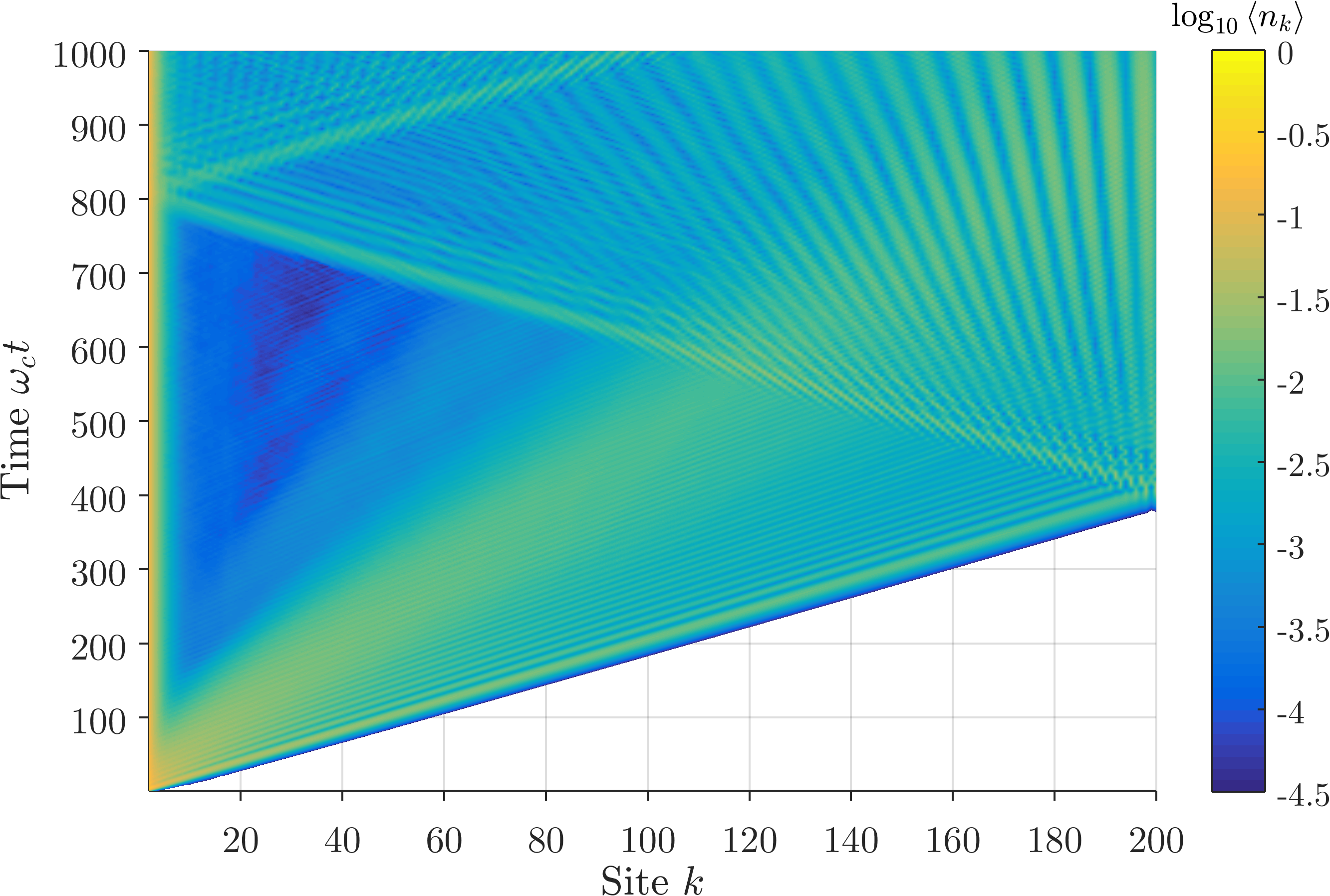}
	\caption{\label{fig:si-results-chain}(color online). The occupation of the chain for $s=1, \alpha =0.2$ shows a clear constant offset on the first few sites indicating the dressing and renormalization of the spin. Additionally the chain exhibits reflections from the end at $\omega_c t\approx 380$ leading to artifacts in the inverse chain mapping. The open system dynamics are correct until the arrival of the reflected waves at $\omega_c t\approx 760$.}
\end{figure}

Instead, we note that this timescale maybe additionally modified by polaronic dressing (spin-bath entanglement), which leads to a renormalization (suppression) of the bare TLS energy gap $\Delta$ that drives Ohmic and sub-Ohmic ground states towards their QPTs.
The environment spectrum in Fig. \ref{fig:results-bath-omegak}(a) clearly resolves the emergence of the renormalized TLS's energy gap $\Delta_{r}$.
This ultrafast process, dominated by high frequency ($\omega_{k}>\Delta$) modes, is generally hard to observe but important in organic exciton transfer and has been seen in inorganic semiconductors \cite{hase2003birth}.
According to Silbey and Harris' variational polaron theory for the ground state of the Ohmic SBM \cite{silbey1984variational}, $\Delta_r = \Delta \left( \frac{\Delta}{\omega_c} \right)^{\frac{\alpha}{1-\alpha}}$, which agrees with the peak position extracted from the environmental spectra.
This is shown in Fig. \ref{fig:results-bath-omegak}(b) for $s=1$ and $s=3$ in the regime $0<\alpha\leq 1$.
In the chain basis the renormalization is accompanied by the persistent excitation of the sites closest to the system, which defines an effectively screened system 'seen' by the rest of the environment (see Fig. \ref{fig:si-results-chain}) - an observation familiar in NRG studies of the related Kondo problem \cite{bulla2008numerical}.
The dynamics of the collective coordinate of the first chain site were also used to verify a novel coherence pumping mechanism in a recent study of a simple photosynthetic pigment-protein complex \cite{chin2013role}.

Accompanying the emission of phonons into the resonant modes, we also observe prominent, damped oscillations for all modes, which vanish on the timescale of spin relaxation (Fig. \ref{fig:results-bath-omegak}).
Recently, Bera \textit{et al.} have shown that strong coupling induces \emph{intermode} entanglement which could induce apparent mixing and damping of individual modes \cite{bera2014generalized,bera2014stabilizing}, but this is predicted for slow and resonant modes, whereas we find damped oscillations at all frequencies.
Given the unitary, energy preserving evolution of the many-body state, this observation warrants further consideration.
To investigate this, we perform a type of state-selective coherent spectroscopy to look at further details within the many-body state.
Fig. \ref{fig:results-bath-omegak}(c) shows the displacement $f_{k}$ of modes projected onto the up spin state, i.e. we compute $f_k^{\uparrow/\downarrow} = \braket{\frac{\openone \pm \sigma_z}{2} \frac{b_k^\dagger+b_k}{2}}$.
It is immediately seen that the displacements do not show \emph{any} damping and oscillate at their natural frequencies over the length of the simulation. 

\begin{figure}
	\includegraphics[width=\linewidth]{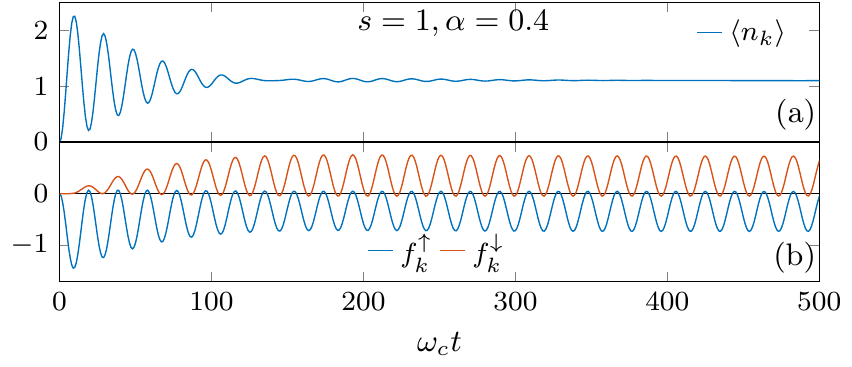}
	\caption{\label{fig:results-polaron-updown}(color online). (a) The occupation $\braket{n_k}$ and (b) the spin-projected displacements $f_k^{\uparrow/\downarrow}$ of the mode $\omega_k=0.325$ with initial product state.}
\end{figure}

\begin{figure}
	\includegraphics[width=\columnwidth]{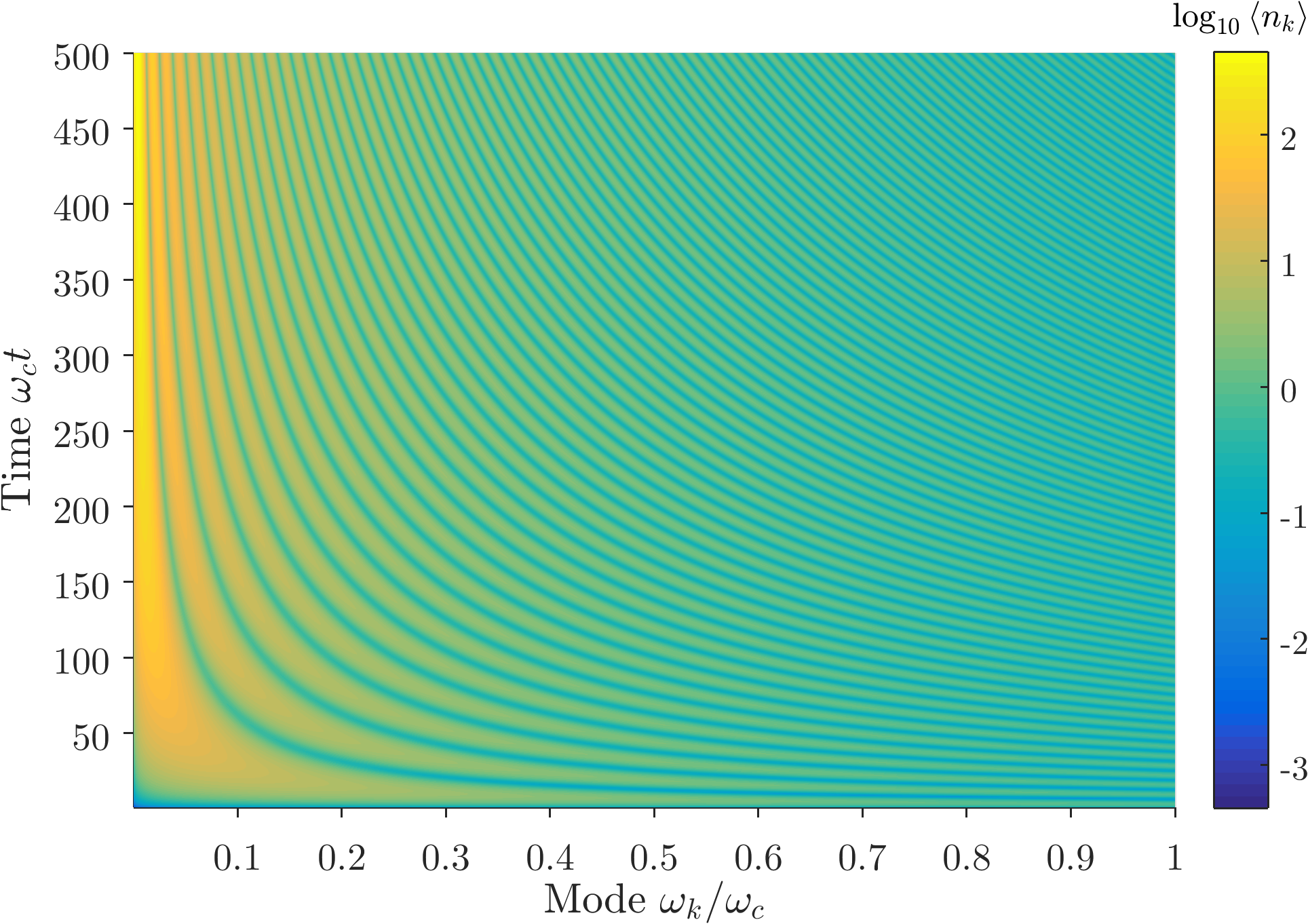}
	\caption{\label{fig:si-results-star}(color online). The phonon occupation for $s=0.75, \alpha=0.4$ with initial product state oscillates even after the relaxation of the spin oscillations at $\omega_c t\approx 150$ due to the finite $\braket{\sigma_z} \neq 0$.}
\end{figure}

Figure \ref{fig:results-polaron-updown} shows the typical dynamics of a high frequency mode.
The positive and negative displacements for $|\uparrow\rangle$ and $|\downarrow\rangle$ are characteristic of polaronic entanglement between the spin and oscillator.
However, it can be seen that their motion is \emph{highly correlated}; maximum mode displacement on one spin state is always obtained at the minimum of the other.
Moreover, the momentum of the mode in each spin state is the same and \emph{preserved}; it is not randomized by the dissipative spin-flip dynamics.
This classical correlation within the entangled state is important, and explains \emph{both} the appearance of a stationary renormalized $\Delta_{r}$ and the apparent relaxation of the mode populations $\langle b_{k}^{\dagger}b_{k}\rangle$.
To motivate this, consider the trial wavefunction $\ket{\Psi(t)} = C_{\uparrow}(t)\ket{\uparrow}\ket{f^\uparrow_k(t)} + C_{\downarrow}(t)\ket{\downarrow}\ket{f^\downarrow_k(t)}$, where we neglect the other environmental modes.
The oscillator wavefunctions for each spin state are time-dependent coherent states $\ket{f^\uparrow_k(t)}=e^{(f_{k}^\uparrow (t)b_k^{\dagger}-f_{k}^{\uparrow *}(t)b_k)}\ket{0}$. 
The expectation value $\braket{\sigma_{x}}$ is then $\braket{ \sigma_{x}}=2\Re[C_{\uparrow}(t)C_{\downarrow}^{*}(t)\braket{f^\downarrow_k(t)|f^\uparrow_k(t)}]$.
The oscillatory displacements we find fit $2f_{k}^{\uparrow}(t)\approx-g_{k}\omega_{k}^{-1}(1-e^{i\omega_{k}t})$ and $2f_{k}^{\downarrow}(t)\approx g_{k}\omega_{k}^{-1}(1+e^{i\omega_{k}t)})$, after about one oscillation period.
As $\langle \sigma_{x}\rangle$ is determined by the overlap of the oscillator wavefunctions entangled with each spin state, we see that the relative displacement of these wavefunctions is preserved for \emph{all} further times, obeying  $f_{k}^{\downarrow}-f_{k}^{\uparrow}=g_{k}\omega_{k}^{-1} $.
Thus the correlated motion provides a constant renormalization of the spin tunneling, which when summed over all modes leads to the $\Delta_{r}$ predicted by ground state theories and observed here as the peak position of the environmental absorption.
Indeed, had the relative displacement been time-dependent then $\langle \sigma_{x}\rangle$ would not relax to a stationary value. 
Next we see that the mode population $\braket{b_{k}^{\dagger}b_{k}} =  |C_{\uparrow}(t)|^{2}|f_{k}^{\uparrow}(t)|^2+  |C_{\downarrow}(t)|^2|f_{k}^{\downarrow}(t)|^2 =\frac{1}{2}g_{k}\omega_{k}^{-1}(1+\langle \sigma_{z}\rangle \cos(\omega_{k}t))$.
Again, we see the correlated motion only leads to oscillations of the population when the spin is out of equilibrium ($\langle \sigma_{x}\rangle\neq1)$, and explains why the apparent population damping for all high frequency modes is set by the relaxation of the spin.
Interestingly, this analysis predicts persistent population oscillations for $\alpha>\alpha_{c}$ in sub-Ohmic baths, or a biased TLS, which is confirmed in Fig. \ref{fig:si-results-star}.
However, we note that in real systems the environment must be connected, albeit weakly, to an external bath, and oscillations will vanish at very long times (\emph{all} observables become stationary).

These results show how broadband environmental dynamics can be analyzed at their various significant frequency scales.
Indeed, we have only considered a small subset of the SBM physics; similar analysis must also be applied to resonant and slow frequency modes, the dynamical hierarchy and feedback between their contributions and the role of any isolated modes, to build the full picture.
This dynamical dissection, made possible by our VMPS many-body approach, not only offers new insights into the fundamental aspects of complex open systems, it may also guide the construction of cheaper \emph{ans\"atze} for describing their ground states and dynamics \cite{bera2014generalized, bera2014stabilizing, Yao2013, Zhang2010, iles2014environmental}.
More immediately, the observation of \emph{conserved} vibrational coherences during heavily damped spin relaxation has relevance for time-resolved observation of transfer, activation and deactivation of vibrational coherences, recently shown to be a powerful tool for exploring ultrafast, coherent processes in optoelectronic systems and biological photoreactions \cite{bakulin2015vibrational, musser2015evidence, liebel2013broad, liebel2014direct}.
Finally, we note that further extensions are required for general applications, many of which - such as finite temperatures and damping of the primary environment - have already been developed in related methods, such as the DMRG-based TEDOPA algorithm of Prior \textit{et al.} \cite{prior2010efficient}.

\begin{acknowledgments}
	We thank R. H. Friend for making this work possible. F. A. Y. N. S. and A. W. C. gratefully acknowledge the support of the Winton Programme for the Physics of Sustainability and EPSRC.
\end{acknowledgments}

\bibliography{TDVMPS}

\end{document}